\documentclass[fleqn,twoside]{article}
\usepackage{latexsym}
\usepackage{espcrc2}
\usepackage{graphicx}
\usepackage[figuresright]{rotating}

\newcommand{\AmS}{{\protect\the\textfont2
  A\kern-.1667em\lower.5ex\hbox{M}\kern-.125emS}}
\newcommand{\ba}{\begin{eqnarray}}
\newcommand{\ea}{\end{eqnarray}}
\newcommand{\be}{\begin{equation}}
\newcommand{\ee}{\end{equation}}

\def\OMIT#1{{}}
\def\eqn#1{Eq.~\ref{#1}}

\def\fig#1{Figure~\ref{#1}}
\def\figs#1#2{Figures~\ref{#1}--\ref{#2}}
\def\tab#1{Table~\ref{#1}}

\def\ap{\alpha_P}
\def\av{\alpha_V}
\def\as{\alpha_s}
\def\ms{\overline{\rm MS}}
\def\ams{\alpha_{\ms}}

\def\etal{et~al.}
\def\cc{\bar{c}c}
\def\bb{\bar{b}b}
\def\tr{{\rm Tr\,}}

\def\vereq#1#2{\lower3pt\vbox{\baselineskip1pt\lineskip1pt
     \ialign{\\$#1\hfill##\hfil\\$\crcr#2\crcr\sim\crcr}}}

\makeatletter
\def\fmslash{\@ifnextchar[{\fmsl@sh}{\fmsl@sh[0mu]}}
\def\fmsl@sh[#1]#2{%
  \mathchoice
    {\@fmsl@sh\displaystyle{#1}{#2}}%
    {\@fmsl@sh\textstyle{#1}{#2}}%
    {\@fmsl@sh\scriptstyle{#1}{#2}}%
    {\@fmsl@sh\scriptscriptstyle{#1}{#2}}}
\def\@fmsl@sh#1#2#3{\m@th\ooalign{$\hfil#1\mkern#2/\hfil$\crcr$#1#3$}}

\title{Charmonium with three flavors of dynamical quarks\thanks{
Talk and poster presented by P. Mackenzie and D. Menscher.}}

\author{Massimo~di~Pierro\address[FNAL]{
Fermi National Accelerator Laboratory, P.O. Box 500, Batavia, IL 60510},
Aida~X.~El-Khadra\addressmark\address[UIUC]{
Department of Physics, University of Illinois, Urbana, IL 61801},
Steven~Gottlieb\addressmark[FNAL]\address[IU]{
Department of Physics, Indiana University, Bloomington, IN 47405},
Andreas~S.~Kronfeld\addressmark[FNAL], Paul~B.~Mackenzie\addressmark[FNAL],
Damian~P.~Menscher\addressmark[UIUC], Mehmet~B.~Oktay\addressmark[UIUC],
and James~N.~Simone\addressmark[FNAL]
}
       
\begin{document}

\begin{abstract}
We present a calculation of the charmonium spectrum with three flavors
of dynamical staggered quarks from gauge configurations that were 
generated by the MILC collaboration. We use the Fermilab action for the
valence charm quarks. Our calculation of the spin-averaged 1P--1S and
2S--1S splittings yields a determination of the strong coupling, 
with $\ams(M_Z) = 0.119 (4)$. 
\vspace{1pc}
\end{abstract}

\maketitle

\section{INTRODUCTION}

The current experimental program of precision flavor physics at the
$B$ factories, at CESR-c, and at the Tevatron needs accurate lattice
QCD calculations of the relevant hadronic matrix elements to yield
stringent constraints on the CKM sector of the standard model. 
Precision lattice QCD results in turn require that the systematic errors 
associated with lattice calculations be brought under control to the 
desired accuracy. The most important sources of systematic error in 
lattice calculations include the incomplete inclusion of sea quarks 
(quenched approximation), lattice spacing artifacts, perturbative errors, 
and the chiral extrapolation. 
We are planning a series of lattice calculations of the 
phenomenologically most important quantities in the $B$, $D$, $\cc$
and $\bb$ systems. We will address the first two sources of systematic
error by performing simulations with highly improved actions on gauge 
configurations with $n_f=2+1$ improved staggered quarks \cite{stagimp}. 
This effort must be complemented by the corresponding perturbative matching 
calculations of the improvement coefficients and currrent 
renormalizations. This should be possible with recent advances
in automated perturbation theory \cite{trottier}.

The MILC collaboration has generated dynamical gauge configurations 
\cite{milc} using improved staggered and gluon actions. Their 
configurations include three (or $2+1$) flavors of light staggered 
fermions at several different light quark masses ranging from $m_s$ to 
$m_s/5$. Hence, the systematic errors usually associated with the 
quenched approximation should be absent with these configurations. 
Furthermore, since numerical simulations with rather light quark masses
are feasible with staggered actions, the issue of chiral extrapolations
may be carefully studied.

The heavy quark action used in this work is based on Ref.~\cite{kkm}.
It is related to NRQCD, but uses the four component fields 
and operators of the Wilson action rather than the two component fields 
and operators of NRQCD. Similar to NRQCD, the space-like and time-like 
components of the operators are uncoupled, and the coefficients of the
operators are mass dependent. This action smoothly interpolates between 
an ordinary light quark action as $am\rightarrow 0$ and NRQCD when
$am>1$, but is applicable at all values of $am$. 
Our formalism can be regarded as a summation of terms of the 
form $(am)^p$ to all orders in the normalizations of operators,
which is useful when $m\gg\Lambda_{\rm QCD}$.
To $O(a)$, our action uses the same operators as the clover action
\cite{sw}.  Starting at $O(a^2)$, the operators are somewhat
different from those in the analogous light quark action. 
The reason is that a two-hop correction to the Wilson time derivative 
operator cannot be used because it introduces ghost states for heavy 
quarks. Its effects must instead be duplicated with Hamiltonian-style 
operators \cite{bo}.
The action may be particularly useful for the charm quark 
on lattice spacings with $am_c<1$.
It has smaller discretization errors than standard light quark actions
since the $(am_c)^p$ errors, 
which it resums and eliminates, are much larger than
the remaining $(a \Lambda_{\rm QCD})^p$ discretization errors.
Since it has a well-defined $a\rightarrow 0$ limit, it may be more
convergent for $am<1$ than NRQCD, which does not.

Now that we are calculating with dynamical quarks, lattice spacings
obtained from the simplest quantities should all agree within errors.
We start our work with a study of the charmonium spectrum, to test
our methods and lattices. As a byproduct, we obtain a new determination 
of the strong coupling. 

\section{DETAILS OF THE CALCULATION}

MILC uses the Asqtad action \cite{milc} for the staggered fermions which 
contains errors of $O(\as a^2)$, and an improved 
gluon action with $O(\as^2 a^2)$ errors. For the charm quarks we use the 
Sheikholeslami-Wohlert action \cite{sw} (which has $O(\as a)$ errors) 
with the Fermilab interpretation \cite{kkm}.
The quenched gauge configurations are generated with the Wilson gauge 
action. \tab{params} lists the simulation parameters for 
all three lattices. The quenched lattice has a slightly smaller lattice 
spacing than the MILC lattices.
\begin{table}[htb]
\caption{Simulation parameters for the three lattices.}
\begin{tabular}{llll}
\hline
\hline
Size		& $16^3\times32$ & $20^3\times64$ & $20^3\times64$ \\
$n_f$		& 0	& 3 & $2+1$ \\
$\beta$		& 5.9 	& 6.85	& 6.76	\\
configs		& 300 & 174 & 298 \\
$am_s$		& $\infty$ & 0.05 & 0.05 \\
$am_l$		& $\infty$ & 0.05 & 0.01 \\
$\kappa_{\rm ch}$ & 0.1227 & 0.113 & 0.113 \\
wf's & $\delta$, 1S, 2S & $\delta$, 1S & $\delta$, 1S \\
\hline
\hline
\end{tabular}
\label{params}
\end{table}

\begin{figure}[htb]
\includegraphics*[width=7.5cm]{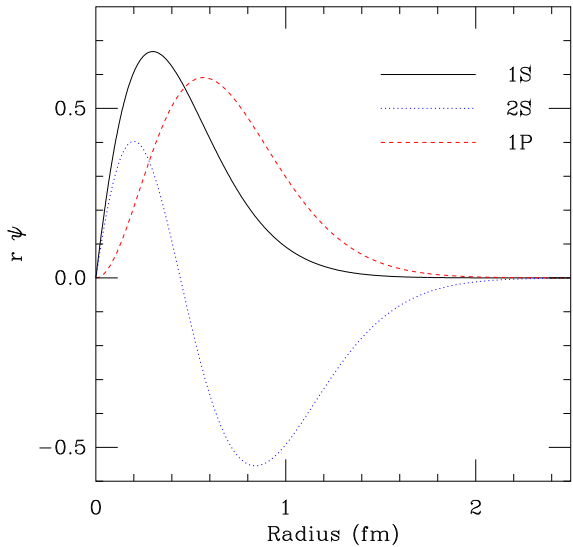}
\caption{Wavefunctions for the $1S$, $2S$, and $1P$ states in
charmonium.}
\label{wf}
\end{figure}

As usual, we calculate charmonium two-point functions using smeared
source and sink operators.  For this purpose, when working on MILC
lattices, we use the Richardson 
potential \cite{rich} model wave functions shown in \fig{wf}.
The quenched propagators were generated using
Coulomb wave functions.

\subsection{Fits}

As shown in Refs.~\cite{lepage,morning}, constrained fits allow
for better control of the systematic error due to excited
state contributions, because they allow us to fit the correlators
to a large number of states without loss of accuracy. Furthermore,
with constrained fits one is also able to use all of the time slices 
in the fits without adjusting $t_\mathrm{min}$ and $t_\mathrm{max}$ to 
determine the best fit. We investigate this issue by comparing 
constrained and unconstrained fits. We fit meson correlators to the form
\ba \label{GtE}
\lefteqn{G(t; \{Z_n\}, \{E_n\}) = } \\ \nonumber
 & & \;\;\;\;\; \sum_{n} Z_n^2 (e^{-E_n t} + e^{-E_n(T-t)}) \,.
\ea 
We force our energy levels to be ordered by defining
\be \label{Eorder}
\Delta E_n = E_n - E_{n-1} \equiv \exp(\epsilon_n) \;.
\ee
In our fits we use Baysian statistics,
\be  \label{bayes}
\chi^2 \rightarrow \chi_\mathrm{aug}^2\equiv\chi^2+\chi_\mathrm{prior}^2 \;,
\ee
where $\chi_\mathrm{prior}^2$ is used to constrain $E_0$, $\epsilon_n$,
and $Z_n$ to a predetermined range. 
\OMIT{
The constraints allow us to fit to
all timeslices $>$ 0, and to more energy levels than would be possible with
unconstrained fits.  
We exclude timeslice 0 because the contact term
simply adds an arbitrary delta function to the fit.}

\begin{figure}[htb]
\includegraphics*[width=7.5cm]{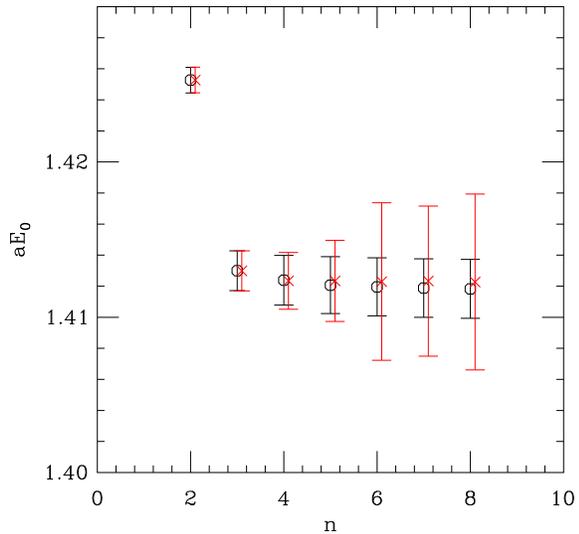}
\caption{Comparison of fit results from constrained and unconstrained fits.
Shown is the $\eta_c$ ground state energy---obtained from the local-local
correlator as a function of the number of states in the fit ($n$).
 $\circ$: constrained fit, $\times$: unconstrained fit}
\label{E0vsn}
\end{figure}

For simplicity, we test our fit program on the quenched lattice.
\figs{E0vsn}{E1vsn} show our fit results for the $\eta_c$ ground and
first excited state energies as functions of the number of states in 
the fit ($n$) obtained from the $\eta_c$ propagator with a 
$\delta$-function source and sink. 
\begin{figure}[htb]
\includegraphics*[width=7.5cm]{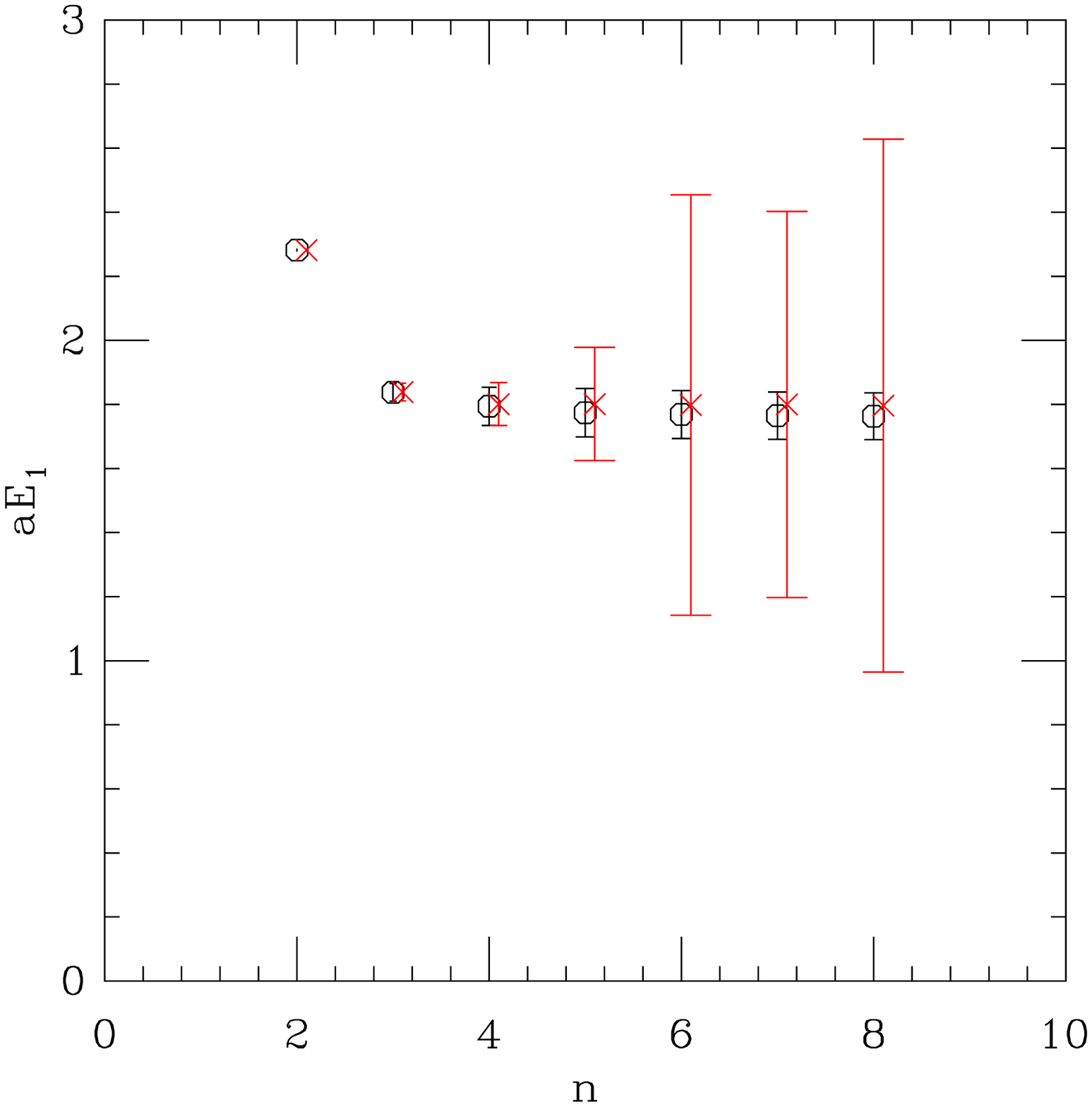}
\caption{Same as \fig{E0vsn}, but for the first excited state,
the $\eta_c'$.}
\label{E1vsn}
\end{figure}
We observe that the fit results --- particularly the error bars --- 
for the ground and first excited 
state energies are stable under adding more states to the fit when 
constrained fits are used, but not in the case of unconstrained fits.
The unconstrained fits do not use \eqn{bayes}, but they do use the
energy ordering constraint of \eqn{Eorder}, which is probably the
reason for the relative stability of the central fit values even in
the unconstrained case.

\begin{figure}[htb]
\includegraphics*[width=7.5cm]{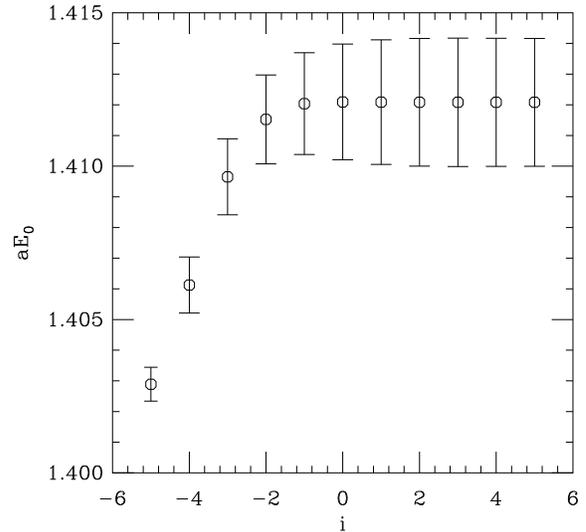}
\caption{Variation of the $\eta_c$ ground state energy fit result with 
the prior width. $i$ is defined via $\sigma = 2^i \sigma_0$, where $\sigma$
and $\sigma_0$ are the varied and standard choice of prior widths 
respectively. }
\label{priors}
\end{figure}

The prior constraints must be chosen so that they do not unduly influence
the physical results. \fig{priors} shows the dependence of the ground
state fit on the prior width. 
The fit includes four states and the ranges for all priors are
varied together with the ground state energy prior width.
We see that the fit results are stable, once the prior width is large 
enough. Furthermore, the error on the ground state energy is unaffected 
by the variation of the prior width, after the plateau is reached.
\tab{prior} lists typical choices for the energy prior values and 
ranges used in our fits.
\begin{table}[htb]
\caption{Typical energy prior values and ranges.}
\begin{tabular}{rlll}
\hline
\hline
$n_f$ \hspace{0.5cm} & 0	& 3 & $2+1$ \\ \hline
$\eta_c$: $E_0$	& $1.4 \pm 0.2$ & $1.9 \pm 0.2$  & $1.9 \pm 0.2$	\\
$\Delta E_n$	& $0.4 \pm 0.2$ & $0.4 \pm 0.2$ & $0.4 \pm 0.2$  	\\
$h_c$: $E_0$	& $1.7 \pm 0.2$ & $2.3 \pm 0.2$  & $2.3 \pm 0.2$	\\
$\Delta E_n$	& $0.4 \pm 0.2$ & $0.5 \pm 0.2$ & $0.4 \pm 0.2$  	\\
\hline
\hline
\end{tabular}
\label{prior}
\end{table}
The prior values for the $Z_n$ are chosen by matching the hadron
propagators evaluated at $t=1$ to \eqn{GtE}; the range is usually 
set to a factor of three of the central value.

The results discussed in the next section are obtained from fits to
multiple correlators, making use of the different source and sink
operators listed in \tab{params}. These fits are generally consistent
with fits to the delta-function correlators, albeit with smaller
statistical errors.

\section{THE SPECTRUM}

\begin{figure}[htb]
\includegraphics*[width=7.5cm]{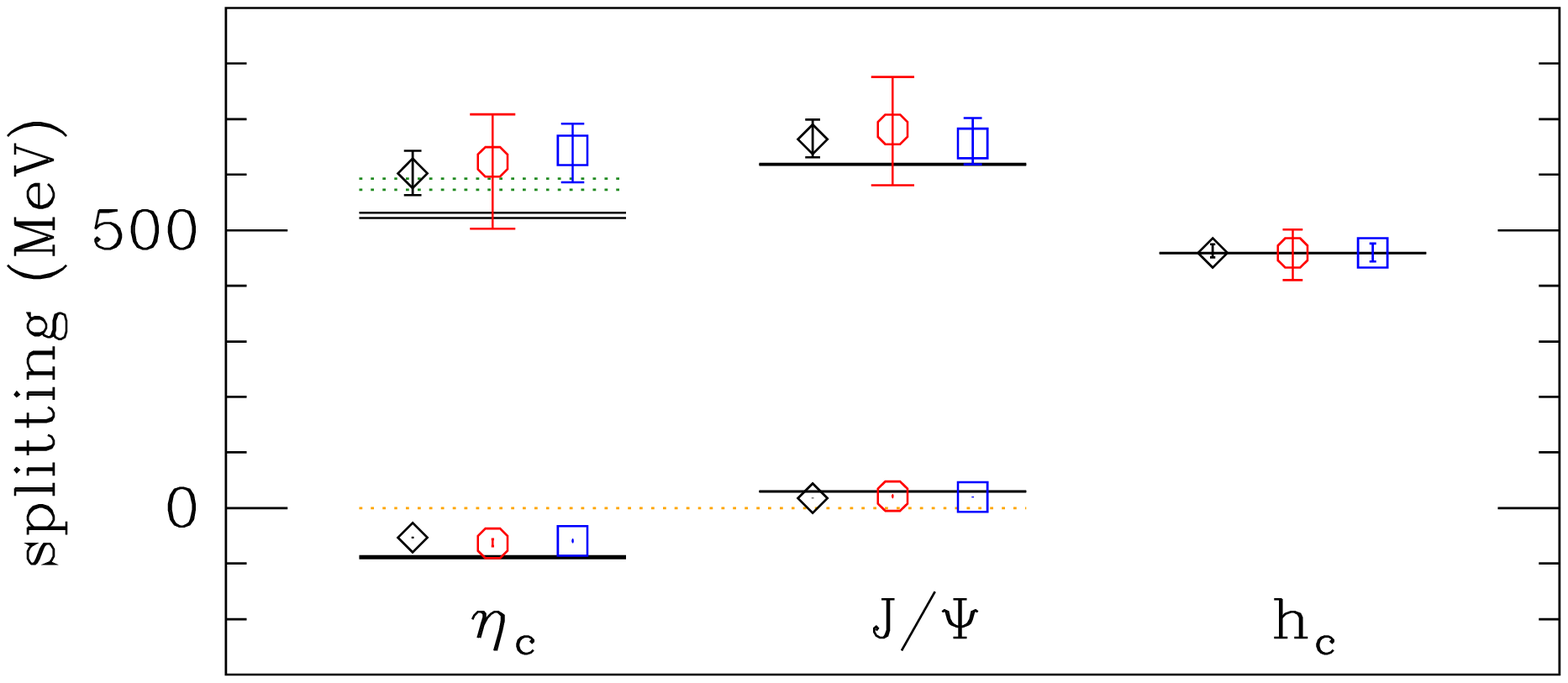}
\caption{The charmonium spectrum in comparison. $\diamond$: $n_f = 0$,
$\circ$: $n_f = 3$, $\Box$: $n_f = 2+1$. }
\label{spectrum}
\end{figure}

\begin{figure}[htb]
\includegraphics*[width=7.5cm]{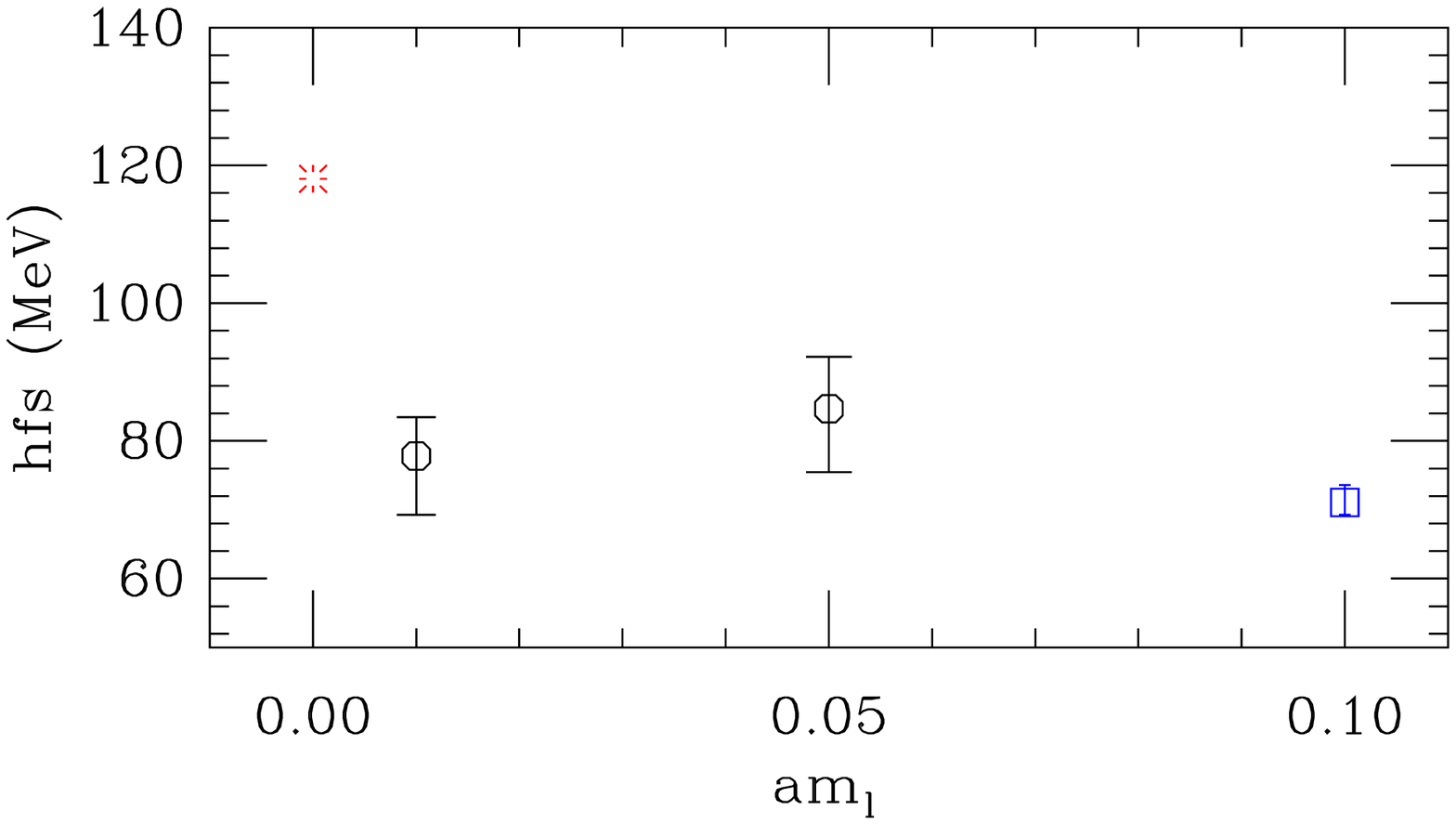}
\caption{The hyperfine splitting {\it vs.}~$am_l$ (circles)). Shown also
is the quenched result (square), positioned in the plot at finite $am_l$ 
for illustration, and the experimental result (burst) positioned at
$am_l = 0$. }
\label{hfs}
\end{figure}

\fig{spectrum} summarizes our results for the charmonium
spectrum. 
Our results for the hyperfine, 1P--1S, and 2S--1S splittings are
shown in \figs{hfs}{2s1s} as functions of the light quark mass.
\begin{figure}[htb]
\includegraphics*[width=7.5cm]{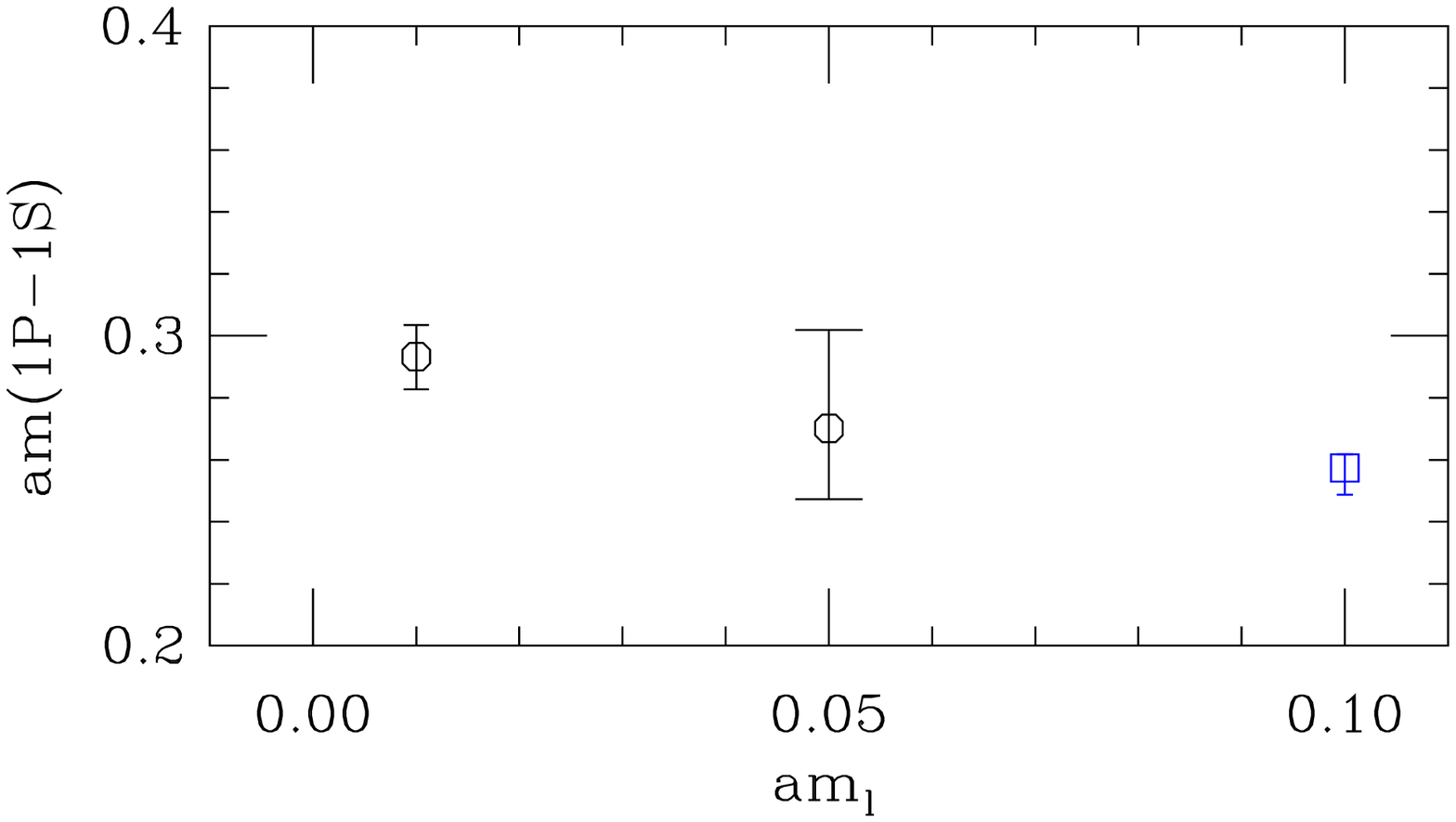}
\caption{The 1P--1S splitting {\it vs.}~$am_l$ (circles). Shown also
is the quenched result (square), positioned in the plot at finite $am_l$ 
for illustration. }
\label{1p1s}
\end{figure}
\begin{figure}[htb]
\includegraphics*[width=7.5cm]{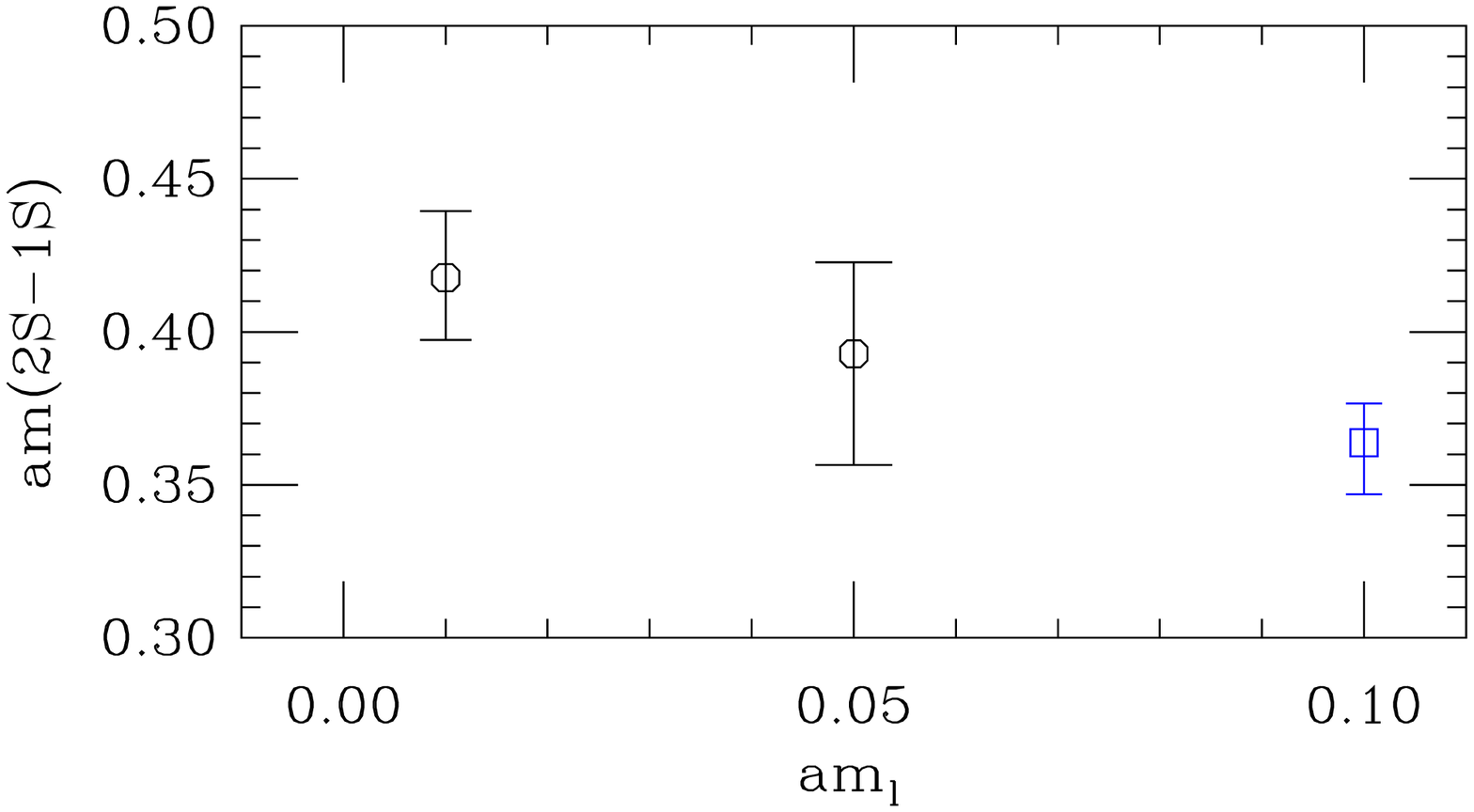}
\caption{The 2S--1S splitting {\it vs.}~$am_l$ (circles). Shown also
is the quenched result (square), positioned in the plot at finite $am_l$ 
for illustration. }
\label{2s1s}
\end{figure}
We observe very little light quark mass dependence in the hyperfine
splitting, and our chirally extrapolated result still disagrees with 
experiment, although the inclusion of dynamical quarks has removed one
possible cause of a small hyperfine splitting: the fact that
the short distance coupling constant is too small in the quenched
approximation.
We note that the charm quark action is only $O(a)$ improved.
The leading order operator which controls the spin splitting is 
$\bar{\psi} \sigma_{\mu \nu}F_{\mu \nu}\psi$. Its coefficient
is being included only at tadpole improved tree-level, and it is
plausible that the observed discrepancy is a result of both 
$O(\as a)$ errors and $O(a^2)$ lattice spacing artifacts. We will 
be able to study this issue further once our $O(a^2)$ improved 
action \cite{bo} is ready for numerical simulations.

The spin-averaged 1P--1S and 2S--1S splittings are considerably 
less sensitive to the leading order lattice artifacts; they are 
used to determine the lattice spacing. \figs{1p1s}{2s1s} indicate 
that the dependence of the spin-averaged splittings on the light 
quark mass is mild.

The extent to which the lattice spacings from the 1P--1S and 2S--1S 
splittings disagree with each other is an indication of residual 
systematic errors in our simulation, which are a combination
of higher order lattice spacing and sea quark effects. 
For this comparison we compute the ratio of lattice spacings,
\be
R \equiv \frac{\frac{\Delta M ({\rm 2S-1S})}{\Delta M ({\rm 1P-1S})}^{\rm lat}}
         { \frac{\Delta M ({\rm 2S-1S})}{\Delta M ({\rm 1P-1S})}^{\rm exp}}
\ee
\tab{ainv} compares the lattice spacings obtained on all three
\begin{table}[b]
\caption{The lattice spacings from the spin-averaged 1P--1S and 2S--1S 
splittings for the three lattices with statistical error bars.}
\begin{tabular}{cccc}
\hline
\hline
$n_f$	& $a^{-1}$(1P--1S) 	& $a^{-1}$(2S--1S) & $R$ \\ 
  	& (GeV)				&  (GeV) &  \\ \hline
 0 	& $1.783^{+0.062}_{-0.030}$ 	& $1.637^{+0.079}_{-0.059}$ &
	$1.089^{+0.057}_{-0.052}$ \\
 3 	& $1.70^{+0.16}_{-0.18}$ 	& $1.52^{+0.15}_{-0.11}$ &
	$1.12^{+0.14}_{-0.17}$ \\
$2+1$ 	& $1.564^{+0.058}_{-0.053}$ 	& $1.426^{+0.073}_{-0.070}$ &
	$1.097^{+0.073}_{-0.068}$ \\
\hline
\hline
\end{tabular}
\label{ainv}
\end{table}
lattices. 
On the quenched lattice the deviation of $R$ from unity is about
$1.5$ standard deviations. On the two MILC lattices, the deviation of
$R$ from unity is less significant, because of the still somewhat
large statistical errors. To clarify the situation we need to reduce 
the statistical errors of the results on the MILC lattices.

Now that dynamical fermions are included in the calculations, lattice
spacings from the best-understood quantities should be consistent.
We note that although the lattice spacings obtained from the 1P--1S and the
2S--1S  splittings are consistent with each other, the lattice spacing from
the 1P--1S splitting is consistent with several determinations of the lattice
spacing from the $\Upsilon$ system \cite{gray02}, 
while the one obtained from the 2S--1S splitting
is not.  It is possible that higher order discretization effects are 
responsible for this.
A more interesting possibility is that due to the fact that the 2S is so close
to the $D\overline{D}$ threshold, physical effects from the coupling to
$D\overline{D}$ channels may have a more dramatic effect in the 2S than 
in other states.

\section{THE STRONG COUPLING}

The spin-averaged splittings discussed in the previous section 
are used to determine the strong coupling $\as$. 
The $2+1$ flavor lattices are our most realistic.
We take the 1P--1S splitting as our most reliable determination of the lattice
spacing because of the possible threshold effects in the
2S state.
Following the procedure of Ref.~\cite{lm}, we obtain the strong
coupling from the plaquette. For our
actions we have \cite{nrqcd02}
\ba
 \lefteqn{- \ln \langle \tr U_P \rangle = } \\ \nonumber
 & & 3.0682 \, \ap(q^*) \left[1 - \ap (0.770 + 0.09681 \, n_f) \right]
\ea
where $q^* = 3.33/a$ is the BLM scale for $n_f = 3$. The coupling 
$\ap$ is defined to coincide through one-loop order with $\av$, the 
coupling defined from the heavy quark potential.
Using 
\be
\ams(q)=\ap\left(e^{5/6} \,q\right)\left[1+\frac{2}{\pi}\ap+O(\ap^2)
\right],
\ee
we obtain 
\be
\ams(M_Z)=0.119\pm 0.004.
\ee
The difference between our value and the value reported in Ref.~\cite{nrqcd02}
arises mainly from differing implicit treatment of $O(\alpha^3)$ corrections,
which will soon be known.  The main sources of uncertainty are $O(\alpha^3)$
corrections (3\%), discretization errors (2\%), and statistical errors (1\%).
All three should be significantly reduced soon.

\section{CONCLUSIONS AND OUTLOOK}

We present preliminary results of a calculation of the charmonium
spectrum on gauge configurations generated by the MILC collaboration
using $O(a^2)$ improved actions for the gluons and the $n_f = 2+1$ 
dynamical staggered fermions. We use the $O(a)$ improved clover
action with the Fermilab interpretation for the charm valence
quarks. 
Since the MILC configurations were generated with the correct number
of sea quarks, no extrapolation in $n_f$ is necessary. Furthermore,
the nondegenerate strange and light quark masses in the MILC lattices 
allow us to consider the chiral limit. Comparing results at 
$am_s=am_l =0.05$ and $am_s = 0.05, am_l = 0.01$, we find only mild
light quark mass dependence for all spectral quantities we consider. 
Finally, our calculation yields a new determination of the strong 
coupling where systematic errors due to sea quark effects are
under control. 

For future work, we are planning to improve the statistical accuracy 
of this work. The improvement of the heavy quark action beyond $O(a)$
is in progress \cite{bo}.

\vspace{0.3cm}

We thank the MILC collaboration for the use of their configurations.
We thank Peter Lepage for helpful conversations.
This work was supported in part by the Department of Energy. We thank
the Fermilab Computing Division and the SciDAC program for their support. 
Fermilab is operated by Universities Research Association Inc., under 
contract with the DOE.


\begin{thebibliography}{9}

\bibitem{stagimp} T.~Blum \etal, Phys.~Rev.~D55 (1997) R1133;
J.~Lagae and D.~Sinclair, Phys.~Rev.~D59, (1998) 104511;
G.~P.~Lepage, Phys.~Rev.~D59 (1999) 074502; 
K.~Orginos, D.~Toussaint, and R.~Sugar, Phys.~Rev.~D60 (1999) 054503;
C.~Bernard \etal, Phys.~Rev.~D61 (2000) 111502(R). 
\bibitem{trottier} M.~Nobes and H.~Trottier, these proceedings,
hep-lat/0209017.
\bibitem{milc} C. Bernard \etal (MILC collaboration), 
Phys. Rev. D64 (2001) 054506.
\bibitem{kkm} A. El-Khadra, A. Kronfeld and P. Mackenzie, 
Phys.~Rev.~D55 (1997) 3933.
\bibitem{sw} B. Sheikholeslami and R. Wohlert, Nucl.~Phys.~B259 (1985) 572.
\bibitem{bo} M.~B.~Oktay \etal, these proceedings, hep-lat/0209150.
\bibitem{rich} J.L. Richardson, Phys.~Lett.~B82 (1979) 272.
\bibitem{lepage} G.P. Lepage \etal, Nucl.~Phys.~Proc.~Suppl. 106 (2002) 12.
\bibitem{morning} C. Morningstar, Nucl.~Phys.~Proc.~Suppl. 109 
(2002) 185.
\bibitem{gray02} A.~Gray \etal, these proceedings, hep-lat/0209022.
\bibitem{lm} P. Lepage and P. Mackenzie, Phys.~Rev.~D48 (1993) 2250. 
\bibitem{nrqcd02} C. Davies \etal, these proceedings, hep-lat/0209122.
\end{thebibliography}
\end{document}